\documentclass{aa}  

\usepackage{graphicx}
\usepackage{txfonts}
\usepackage{color}

\begin{document} 

   \title{Investigating the physical properties of galaxies in the Epoch of Reionization with MIRI/JWST spectroscopy}

   %\subtitle{MIRI/JWST spectroscopy of EoR sources}

\titlerunning{MIRI/JWST spectroscopy of EoR sources}

   \author{J.~\'Alvarez-M\'arquez\inst{1}
    \and L. Colina\inst{1,2}
    \and R. Marques-Chaves\inst{1}
    \and D. Ceverino \inst{2,3,4} 
    \and A. Alonso-Herrero \inst{5}
    \and K. Caputi \inst{6,2}
    \and M. Garc\'ia-Mar\'in \inst{7}
    \and A. Labiano \inst{5}
    \and O. Le F\`evre \inst{8}
    \and H. U. Norgaard-Nielsen \inst{9}
    \and G. \"Ostlin \inst{10}
    \and P. G. P\'erez-Gonz\'alez \inst{1}
    \and J. P. Pye \inst{11}
    \and T. V. Tikkanen \inst{11}
    \and P. P. van der Werf \inst{12}
    \and F. Walter \inst{13,14}
    \and G. S. Wright \inst{15}
    }

   \institute{
    $^{1}$Centro de Astrobiolog\'ia (CSIC-INTA), Carretera de Ajalvir, 28850 Torrej\'on de Ardoz, Madrid, Spain; \email{javier.alvarez@cab.inta-csic.es} \\
    $^{2}$Cosmic Dawn Center (DAWN) \\
    $^{3}$Niels Bohr Institute, University of Copenhagen, Lyngbyvej 2, 2100, Copenhagen $\mbox{\normalfont\O}$, Denmark \\
    $^{4}$Universit\"{a}t Heidelberg, Zentrum f\"{u}r Astronomie, Institut f\"{u}rheoretische Astrophysik, Albert-Ueberle-Str. 2, 69120 Heidelberg, Germany\\
    $^{5}$Centro de Astrobiología (CAB, CSIC-INTA), ESAC Campus, E-28692 Villanueva de la Cañada, Madrid, Spain\\
    $^{6}$Kapteyn Astronomical Institute, University of Groningen, P.O. Box 800, 9700AV Groningen, The Netherlands\\
    $^{7}$ European Space Agency, 3700 San Martin Drive, Baltimore, MD21218 \\
    $^{8}$Aix Marseille Universit\'e, CNRS, LAM (Laboratoire d'Astrophysique de Marseille) UMR 7326, 13388, Marseille, France\\
    $^{9}$DTU Space, National Space Institute, Technical University of Denmark, Elektrovej 327, DK-2800 Kgs. Lyngby, Denmark\\
    $^{10}$ Department of Astronomy and Oskar Klein Centre, Stockholm University, SE-10691 Stockholm, Sweden \\
    $^{11}$Department of Physics \& Astronomy, University of Leicester, Leicester, LE1 7RH, U.K.\\
    $^{12}$Leiden Observatory, Leiden University, P.O. Box 9513, NL-2300 RA Leiden, the Netherlands\\
    $^{13}$Max Planck Institute for Astronomy, K\"{o}nigstuhl 17, 69117 Heidelberg, Germany\\
    $^{14}$National Radio Astronomy Observatory, Pete V. Domenici Array Science Center, P.O. Box O, Socorro, NM 87801, USA\\
    $^{15}$UK Astronomy Technology Centre, Royal Observatory, Edinburgh, Black-ford Hill, Edinburgh EH9 3HJ, United Kingdom.
    }
%   \date{15-02-2019}

% \abstract{}{}{}{}{} 
% 5 {} token are mandatory
 
  \abstract
    {The {\it James Webb Space Telescope} (JWST) will provide deep imaging and spectroscopy for sources at redshifts above 6, covering the entire Epoch of Reionization (EoR, 6$< z <$10), and enabling the detailed exploration of the nature of the different sources during the first 1 Gyr of the history of the Universe. The Medium Resolution Spectrograph (MRS) of the mid-IR Instrument (MIRI) will be the only instrument on board JWST able to observe the brightest optical emission lines H$\alpha$ and [OIII]0.5007$\mu$m at redshifts above 7 and 9, respectively, providing key insights into the physical properties of sources during the early phases of the EoR. This paper presents a study of the H$\alpha$ fluxes predicted by state-of-the-art FIRSTLIGHT cosmological simulations for galaxies at redshifts of 6.5 to 10.5, and its detectability with MIRI. Deep (40 ksec) spectroscopic integrations with MRS will be able to detect (S/N $>$ 5) EoR sources at redshifts above 7 with intrinsic star formation rates (SFR) of more than 2 M$_{\odot}$ yr$^{-1}$, and stellar masses above 4-9 $\times$ 10$^7$ M$_{\odot}$. These limits cover the upper end of the SFR and stellar mass distribution at those redshifts, representing $\sim$ 6\% and $\sim$1\% of the predicted FIRSTLIGHT population at the 6.5-7.5 and 7.5-8.5 redshift ranges, respectively. In addition, the paper presents realistic MRS simulated observations of the expected rest-frame optical and near-infrared spectra for some spectroscopically confirmed EoR sources recently detected by ALMA as [OIII]88$\mu$m emitters. The MRS simulated spectra cover a wide range of low metallicities from about 0.2 to 0.02~Z$_{\odot}$,  and different [OIII]88$\mu$m/[OIII]0.5007$\mu$m line ratios. The simulated 10ks MRS spectra show S/N in the range of 5 to 90 for H$\beta$, [OIII]0.4959,0.5007~$\mu$m, H$\alpha$ and HeI1.083$\mu$m emission lines of the currently highest spectroscopically confirmed EoR (lensed) source MACS1149-JD1 at a redshift of 9.11, independent of metallicity.  In addition, deep 40 ksec simulated spectra of the luminous merger candidate B14-65666 at 7.15 shows the MRS capabilities of detecting, or putting strong upper limits on, the weak [NII]0.6584$\mu$m, [SII]0.6717,0.6731$\mu$m, and [SIII]0.9069,0.9532$\mu$m emission lines. These observations will provide the opportunity of deriving accurate metallicities in bright EoR sources using the full range of rest-frame optical emission lines up to 1$\mu$m. In summary, MRS will enable the detailed study of key physical properties such as internal extinction, instantaneous star formation, hardness of the ionizing continuum, and metallicity in bright (intrinsic or lensed) EoR sources.}


   \maketitle
   
%
%-------------------------------------------------------------------
\section{Introduction}\label{Int}

Deep imaging surveys with the {\it Hubble Space Telescope} (HST) have detected galaxies at very high redshifts ($z$ > 5) in large numbers; there are  hundreds of them at photometric redshifts of about 7, and about 200 candidates at redshifts of 8-10,  well within the Epoch of Reionization (EoR) of the universe \citep{Bouwens2015,Oesch2015,Roberts-Borsani2016,Stefanon2017,Oesch2018}.  The combination of {\it HST} and {\it Spitzer} deep imaging has further identified these galaxies as potential strong optical line emitters based on the flux excess in the IRAC 3.6 and 4.5 $\mu$m bands  \citep[e.g.][]{Schaerer2009, Labbe2013, Stark2013, Smit2015, Bouwens2016b, Rasappu2016, Roberts-Borsani2016}. The H$\beta$+[OIII] and the H$\alpha$ lines have large equivalent widths with values of up to (rest-frame) 1000-2000$\AA$ (e.g. \citealt{Faisst2016,Marmol-Queralto2016,Rasappu2016,Smit2016,Caputi2017,Lam2019arXiv}), and a non-linear dependency with redshift (1+$z$)$^\alpha$ with $\alpha$ $\sim$ 1.8 and $\sim$ 1.3 for sources at redshifts $z < 2.5$ and $2.5 < z < 6$, respectively (\citealt{Faisst2016,Marmol-Queralto2016}). 

The spectroscopic confirmation of EoR sources remains very limited. It is based mostly on the Ly$\alpha$ detection (\citealt{Stark2017} for a recent compilation; \citealt{Zitrin2015,Oesch2015b,Jung2019arXiv}), which becomes very inefficient at $z>7$ as only the brightest sources exhibit Ly$\alpha$ emission \citep{Pentericci2011}. Additionally, detection of far-infrared [CII]158$\mu$m and [OIII]88$\mu$m line emitters at redshifts of up to 9.11 have  recently been reported with ALMA \citep{Inoue2016,Carniani2017,Tamura2018,Hashimoto2019,Hashimoto2018b,Smit2018}.

The subarcsec imaging and spectroscopic capabilities of the {\it James Webb Space Telescope} (JWST), combined with its broad spectral range coverage (0.6 to 28 $\mu$m) and its increased  sensitivity of one to two orders of magnitude better than that of previous space observatories such as HST and {\it Spitzer} will provide exquisite data to investigate in detail the physical nature and properties of EoR sources at redshifts above 6. Among the JWST instruments, the Mid-infrared Instrument (MIRI) Medium Resolution Spectrograph (MRS) covering the 5 to 28 $\mu$m spectral range, will be the only instrument capable of detecting the strongest optical lines, H$\alpha$ and [OIII]0.5007$\mu$m  at redshifts well above 6.6 and 9, respectively (\citealt{Wright2015,Wells2015}; and references therein). In addition, while the JWST near-infared spectrograph (NIRSpec) will be covering the UV and blue rest-frame range \citep{Chevallard2019}, the MRS extends the observed spectral range above the   rest-frame [OIII] lines, and well into the 1 $\mu$m  region, where internal extinction is less relevant, and other less explored lines such as [SIII]0.907,0.953 $\mu$m and HeI1.083$\mu$m are present. 

The rest-frame optical and near-IR spectral range covered by the MRS is key to developing a full understanding of the physical properties and mechanisms involved in the earliest stages in the formation of galaxies during EoR. Of the main optical diagnostic lines, the H$\alpha$ is the least affected by internal extinction, and therefore the cleanest tracer of the instantaneous star formation rate (SFR, \citealt{Kennicutt1998} for review) in EoR sources, even if a measurement of internal extinction is not available. Moreover, the combination of H$\alpha$ with Ly$\alpha$ and UV continuum measurements will provide more accurate values for the Ly$\alpha$ and the ionizing escaping fractions. In addition, the ratios [NII]0.6584$\mu$m/H$\alpha$ and [SII]0.6717,0.6731$\mu$m/H$\alpha$  trace the metallicity (Z) in pure star-forming  galaxies (\citealt{Maiolino2019} for a review), and combined with the ratio  [OIII]/H$\beta$  trace the nature of the ionizing source \citep{Kewley2013a}. Several other less explored combinations of line ratios involving the [NII], [SII], and [SIII] lines become available as additional metallicity tracers (\citealt{Maiolino2019} and references therein).  Finally,  the HeI~1.083~$\mu$m, which is the strongest HeI line in the entire optical and  near-IR  range \citep{Porter2005}, can provide in combination with the H$\alpha$ line, a measurement of the hardness of the ionizing continuum, and therefore information on the nature of the ionizing source, as the hydrogen and HeI lines are sensitive to the total amount of photons with energies above 13.6 eV (912$\AA$) and 24.6 eV (504$\AA$), respectively.

Predictions of the nebular spectra of EoR sources from state-of-the-art cosmological simulations \citep{Barrow2017,Ceverino2019,Katz2019} are now available for a direct comparison with future JWST observations.  These simulations follow the physical processes associated with the early formation and evolution of galaxies during the first 1 Gyr of the universe. These simulations predict galaxies in the early universe as strong line emitters, confirmed by the detection of Ly$\alpha$ and, more recently, [OIII]88$\mu$m line emitters at redshifts $\sim$ 7-9. Therefore, the prospects of investigating the nature, evolution, and physical properties of early galaxies with the MIRI spectrograph should be explored in detail.  

This paper presents a study of the detectability of FIRSTLIGHT simulated galaxies at redshifts of 6.5 to 10.5, and realistic MIRI/JWST spectra of the newly discovered high-z [OIII]88$\mu$m emitters detected with ALMA. The paper is structured as follows. The most relevant features of the FIRSTLIGHT simulations and the apparent fluxes of the two strongest optical emission lines ([OIII]0.5007$\mu$m and H$\alpha$) for FIRSLIGHT galaxies at redshifts 6.5 to 10.5 as a function of their SFR, stellar mass (M$_{*}$), and specific star formation (sSFR) are presented in Sect. \ref{Met:FirstLight}, together with a discussion of the detectability of the population of FIRSTLIGHT H$\alpha$ emitters with MRS. Specific examples of MRS simulated spectra for two of the recently detected [OIII]88$\mu$m emitters, MACS1149-JD1 \citep{Zheng2012,Hashimoto2018b} and B14-65666 \citep{Bowler2014,Bowler2017,Hashimoto2019} at a respective  redshift of 9.11 and 7.15, are presented in Sect. \ref{Met:FullSim},  together with the possibilities that MRS opens for the detail studies of their physical properties, such as internal extinction, instantaneous star formation, hardness of ionizing continuum, and metallicity. A summary of the results and future work is presented in Sect. \ref{Conc}. Throughout this paper we use a standard cosmology with matter and dark energy density $\Omega_{\rm m} = 0.3$ and $\Omega_{\Lambda} = 0.7$,  the Hubble constant $H_{0} = 70$ km s$^{-1}$ Mpc$^{-1}$, and the AB magnitude system.

%--------------------------------------------------------------------
\section{FIRSTLIGHT: EoR line emitters from cosmological simulations}\label{Met:FirstLight}

\subsection{FIRSLIGHT overview}

We use the zoom-in cosmological simulations of galaxies of the FIRSTLIGHT project \citep{Ceverino2017,Ceverino2018,Ceverino2019}. Briefly, this consists of a complete mass-selected sample of 289 halos, selected at $z = 5$ in two cosmological boxes of 10 and 20~Mpc~h$^{-1}$ with halo masses between 10$^{9}$ - 10$^{11}$~M$_{\odot}$ (see  details in \citealt{Ceverino2017}). The maximum spatial resolution is 10~pc. The dark matter particle mass resolution is m$_{\rm DM}$~=~10$^{4}$~M$_{\odot}$ and the minimum star particle mass is 100~M$_{\odot}$.

These high-resolution simulations are performed with the ART code \citep{Kravtsov1997,Kravtsov2003,Ceverino2014,CeverinoKlypin2009,Ceverino2019}. They follow the evolution of a gravitating system and the Eulerian gas hydrodynamics, and incorporate other astrophysical processes, such as gas cooling radiation, photoionization heating by the cosmological UV background, a stochastic star formation model, and a model that includes thermal, kinetic and radiative feedback  (see  details in \citealt{Ceverino2017}).

The FIRSTLIGHT database\footnote{Data retrieval is available from \url{http://www.ita.uni-heidelberg.de/~ceverino/FirstLight/index.html}} includes several properties for all snapshots of the main galaxy progenitor of the 289 zoom-in  simulations, such as the virial, stellar and gas masses, and its SFR, in steps of 10~Myr. The database starts when the galaxy reaches the halo mass of M$_{vir} = 10^{9}$~M$_{\odot}$ and ends in the last available snapshot at $z \geq 5$. In general, these galaxies show non-uniform star formation histories, spending most of their time (70\%) in bursts of star formation \citep{Ceverino2018}, consistent with cosmological gas accretion events. In this work, we use all snapshots within the redshift range $6.5 \leq z \leq 10.5$. This sample is composed of 10,064 snapshots, and covers a wide range of stellar masses ($\sim 10^{5-9}$ M$_{\odot}$), SFRs ($\sim 0 - 30$ M$_{\odot}$ yr$^{-1}$), and metallicities ($Z = 3 \times 10^{-5} - 8 \times 10^{-3}$).

In addition to the physical properties mentioned above, spectral energy distributions (SEDs) are also publicly available for all these snapshots \citep{Ceverino2019}. Stellar SEDs are generated using the Binary Population and Spectral Synthesis model \citep[BPASS:][]{Eldridge2017} and assume a \cite{Kroupa2001} initial mass function (IMF). The contribution of nebular emission is also available and assumes the stellar metallicity, and a gas covering factor of one with an electron density of 100 cm$^{-3}$ \citep[see][for details]{Ceverino2019}. 

\subsection{MRS detectability of FIRSTLIGHT EoR line emitters}\label{Res:FirstLight}

\begin{figure}[htb!]
 \centering
$\begin{array}{rl}
    \includegraphics[width=0.48\textwidth]{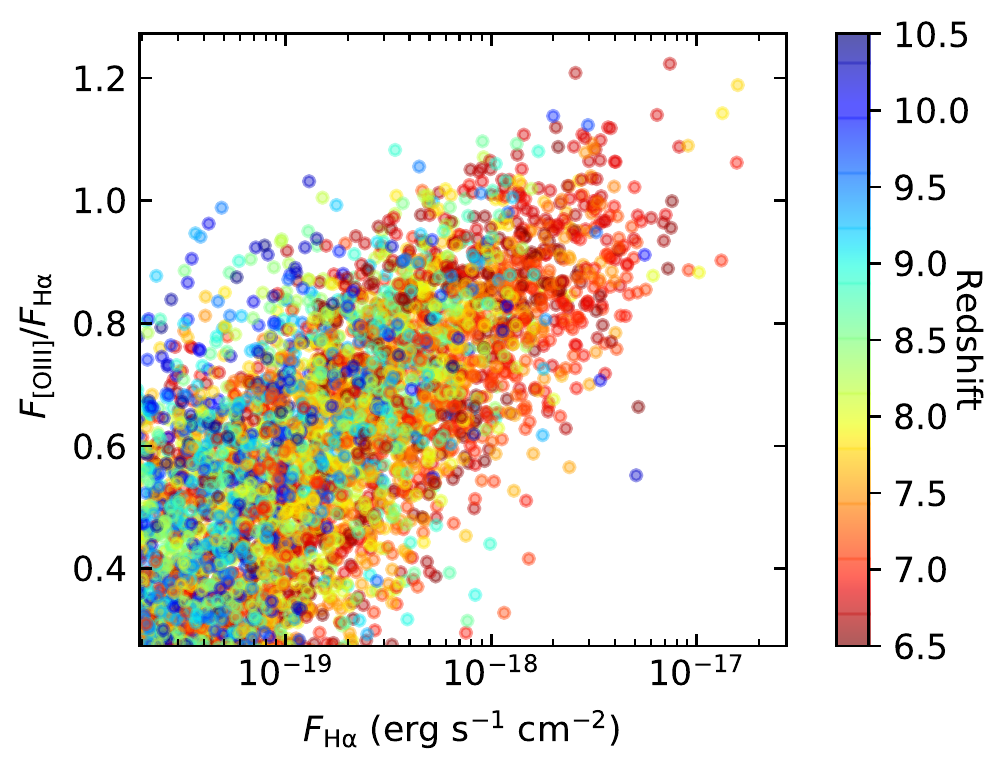}
\end{array}$
\caption{Relation between [OIII]/H$\alpha$ and H$\alpha$ fluxes of the FIRSTLIGHT simulated galaxies at $6.5<z<10.5$. The most luminous galaxies present similar H$\alpha$ and [OIII] fluxes, whereas for fainter galaxies, H$\alpha$ tends to be relatively brighter than [OIII], similar to the values  found in metal-poor, low-z galaxies.}\label{fig:FS_Halpha_[OIII]}
\end{figure}

Luminosities of the two strongest optical emission lines, [OIII]0.5007$\mu$m and H$\alpha$, are extracted for each nebular SED component and converted to observable fluxes (in units of erg~s$^{-1}$~cm$^{-2}$) using the  equation

\begin{equation}
    F_{\rm obs} (\rm [OIII], H\alpha) = \frac{L (\rm [OIII], H\alpha)}{4 \pi D_{L}^2},
\end{equation}

\noindent
where $D_{L}$ is the luminosity distance at a given redshift for the adopted cosmology. 
Figure \ref{fig:FS_Halpha_[OIII]} shows the relation between the ratio   [OIII]/H$\alpha$   the H$\alpha$ emission line fluxes of the simulated galaxies. The most luminous FIRSTLIGHT galaxies present similar H$\alpha$ and [OIII]0.5007$\mu$m fluxes ([OIII]0.5007$\mu$m/H$\alpha$ $\ge$ 1), whereas for fainter galaxies H$\alpha$ tends to be brighter than [OIII]0.5007$\mu$m ([OIII]0.5007$\mu$m/H$\alpha$ < 1), similar to the values  found in metal-deficient, low-$z$ galaxies \citep[e.g.][]{Izotov2011, Hirschauer2016, Izotov2018}. We note, however, that these simulations do not include the effect of dust attenuation, although it is expected to be negligible in low-mass, low-metallicity, high-$z$ galaxies \citep[e.g.][]{Hashimoto2018b}. 

\begin{figure*}[htb!]
 \centering
$\begin{array}{rl}
    \includegraphics[width=1.0\textwidth]{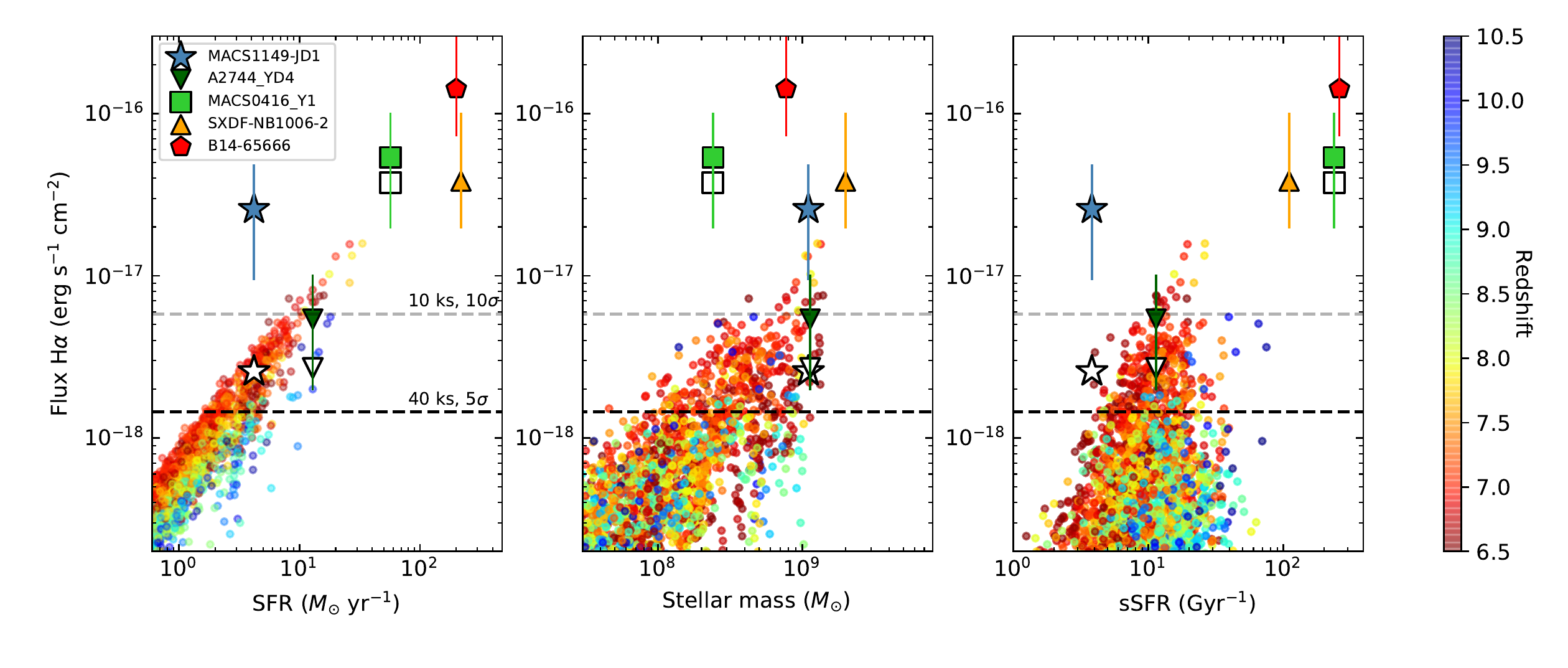}
\end{array}$
\caption{Predicted H$\alpha$ fluxes as a function of star formation rate (left), stellar mass (middle), and specific star formation rate (right) for the FIRSTLIGHT simulated galaxies at $6.5<z<10.5$. Grey and black horizontal dashed lines mark respectively the 10$\sigma$ and $5\sigma$ limits of medium deep (10 ks) and deep (40 ks) MRS spectroscopic observations. Other known $z > 7$ galaxies with detected [OIII]~$88\mu$m emission lines \citep{Inoue2016, Laporte2017, Hashimoto2019, Hashimoto2018b, Tamura2018} are also shown with filled symbols \citep[SFR and stellar mass measurements of A2444$\_$YD4 SXFD-NB10006-2 have been corrected for the lower proportion of low-mass stars in the Kroupa IMF relative to the standard Salpeter assumed in][respectively]{Inoue2016, Laporte2017}. For these galaxies, ratios of [OIII]~$88 \mu$m / [OIII]~$0.5007 \mu$m $ = 10$ and [OIII]~$0.5007 \mu$m / H$\alpha = 1.1$ are used to derive the expected H$\alpha$ fluxes (vertical lines show the minimum and maximum expected H$\alpha$ fluxes considering different line ratios of R[OIII] and [OIII]$0.5007\mu$m/H$\alpha$; see text for details). For the galaxies magnified by gravitational lensing, such as MACS1149-JD1, A2744$\_$YD4, and MACS0416$\_$Y1, the de-magnified expected fluxes are shown as empty symbols (SFR and $M_{*}$ measurements refer to intrinsic values). \label{fig:FS_galaxies}}
\end{figure*}

Figure \ref{fig:FS_galaxies} shows the derived H$\alpha$ fluxes of all snapshots of the main galaxy progenitor of the FIRSTLIGHT simulations with redshifts between 6.5 and 10.5 as a function of their SFRs, stellar masses, and specific SFRs. Overall, these galaxies show a linear relation between H$\alpha$ fluxes and SFRs, as expected, since FIRSTLIGHT SFRs are computed using stellar particles younger than $\sim 10$ Myr that produce copious amounts of ionizing photons that ionize the surrounding gas. The relation between H$\alpha$ flux and stellar mass, and therefore sSFR, is nevertheless much broader due to the stochastic star formation histories in the simulations \citep[for details, see][]{Ceverino2018}. This sample is dominated by numerous low-mass galaxies with extremely faint H$\alpha$ emission for JWST spectroscopy, characterized by median values of $\widetilde{F} (\rm H\alpha) = 3.8$, $2.4$, and $1.5 \times 10^{-20}$~erg~s$^{-1}$~cm$^{-2}$ in the redshift intervals of 6.5-7.5, 7.5-8.5, and $z > 8.5$, respectively. However,  a fraction of the galaxies show much higher fluxes of around $F (\rm H\alpha) \sim 10^{-18}$ - $10^{-17}$~erg~s$^{-1}$~cm$^{-2}$ that are accessible to observation with MIRI/JWST spectroscopy.

In order to study the detectability of H$\alpha$ in such galaxy population, we use the expected MRS limiting sensitivity curves of \cite{Glasse2015}. We note that these sensitivity curves refer to point-like sources with spectrally unresolved lines. By using medium deep (10 ks) and deep (40 ks) on-source MRS spectroscopic observations we find limiting H$\alpha$ fluxes of $ \simeq 5.8 \times 10^{-18}$ erg s$^{-1}$ cm$^{-2}$ ($10\sigma$ in 10~ks) and $\simeq 1.4 \times 10^{-18}$ erg s$^{-1}$ cm$^{-2}$ ($5\sigma$ in 40~ks), respectively.\footnote{The sensitivity is roughly constant within the wavelength range covered by the MRS Channel 1, from 4.9 to 7.6 $\mu$m, i.e. H$\alpha$ redshifted to $6.5 < z < 10.5$ \citep[see][]{Glasse2015}.} As shown in Figure \ref{fig:FS_galaxies}, this means that for the entire $6.5 < z < 10.5$ FIRSTLIGHT sample, only a small fraction of about 6.2, 1.1, and $0.4 \%$ of FIRSTLIGHT galaxies in the redshift range of 6.5-7.5, 7.5-8.5, and $z > 8.5$, respectively, would be detected (S/N $\geq$ 5) in deep 40 ks observations. This indicates that only the most luminous FIRSTLIGHT simulated galaxies, those with star formation rates higher than $1.6$, $1.9$, and $3.9$ M$_{\odot}$ yr$^{-1}$, and stellar masses higher than $ 4$, $9$, and $14 \times 10^{7}$ M$_{\odot}$ in the redshift intervals of 6.5-7.5, 7.5-8.5, and $z > 8.5$, respectively, will be accessible for detailed studies with MRS spectroscopy in a moderate amount of observing time (40 ks). 

It should be noted, however, that the FIRSTLIGHT simulations are limited to halo masses of a few times $10^{11} M_{\odot}$ within a cosmological volume of $\sim2 \times 10^{4}$~Mpc$^{3}$. As shown in the middle panel of Figure \ref{fig:FS_galaxies}, this limits simulated galaxies to have stellar masses above $2 \times 10^9$~M$_{\odot}$. However, massive EoR galaxies have been recently detected by ALMA as [OIII]$88\mu$m emitters \citep{Inoue2016, Hashimoto2019, Tamura2018}. These galaxies are relatively massive, $M_{*} = (2-5) \times 10^{9}$ M$_{\odot}$, and show very strong [O III] $88\mu$m line fluxes, $\simeq (0.6 - 17.5) \times 10^{-18}$ erg s$^{-1}$ cm$^{-2}$. The UV-luminous, high EW[H$\beta$+OIII], Ly$\alpha$-emitter sources \citep{Roberts-Borsani2016,Stark2017}, could also belong to the same class of EoRs. Assuming a wide range of [OIII]0.5007$\mu$m/[OIII]88$\mu$m (hereafter R[OIII]) and [OIII]$0.5007\mu$m/H$\alpha$ line ratios (R[OIII]$= 6.5 - 10$ and [OIII]$0.5007\mu$m/H$\alpha = 0.59 - 1.93$, see Table \ref{tab_MRS_sim}), these galaxies will show H$\alpha$ fluxes of about $(0.2 - 30) \times 10^{-17}$ erg s$^{-1}$ cm$^{-2}$ (see Figure \ref{fig:FS_galaxies}), well above the detection limits even for medium-deep (10 ks) MRS observations. On the other hand, for more typical, less luminous galaxies, the power of strong gravitational lensing may add the required boost in the apparent fluxes necessary to reach the MRS sensitivity. Therefore, high S/N optical ($\sim$ 0.5 to 1 $\mu$m)  emission line spectra will become available with MRS for the first time at such early cosmic times, providing the opportunity of characterizing several of the physical properties of these sources. An exploration of these possibilities is presented in the following section with two specific examples.  

\section{MIRI/JWST spectroscopy: EoR [OIII]88$\mu$m line emitters}\label{Met:FullSim}

In the previous section we conclude that all ALMA detected [OIII]88$\mu$m sources, and also known UV-luminous Lyman-alpha emitters (LAEs) \citep{Stark2017}, in the EoR will be easily studied using the H$\alpha$ emission line with a medium-deep (10ks) and deep (40ks) MRS observations. In the following we present realistic MRS simulated observations of the rest-frame optical and near-IR spectrum ($\sim$0.5 - 1.2 $\mu$m) for two recently ALMA detected [OIII]88$\mu$m emitters, MACS1149-JD1 \citep{Zheng2012,Hashimoto2018b} and B14-65666 \citep{Bowler2014,Bowler2017,Hashimoto2019}. MACS1149-JD1 is a lensed galaxy with a magnification factor of $\sim$10 at a redshift of 9.11, being the highest-$z$ spectroscopically confirmed galaxy based on an emission line. Its derived intrinsic SFR of 4.2M$_{\odot}$ yr$^{-1}$, sSFR of 4 Gyr$^{-1}$, stellar mass of 1.1$\times$ 10$^9$ M$_{\odot}$, and observed [OIII]~88$\mu$m flux of $3 \times 10^{-18}$~erg~s$^{-1}$~cm$^{-2}$ places it within the range of H$\alpha$ fluxes clearly detectable with medium-deep   MRS spectroscopy. On the other hand, B14-65666 is a Lyman-break galaxy system of two sources likely interacting or merging at redshift 7.15, identified as UV bright with an absolute magnitude of M$_{UV} \sim -22.3$, which  places it in the range of luminous LAEs \citep{Roberts-Borsani2016,Stark2017}. The global system has a derived SFR of 200 M$_{\odot}$ yr$^{-1}$, sSFR of 259 Gyr$^{-1}$, stellar mass of 7.7 $\times$ 10$^8$ M$_{\odot}$, and [OIII]~88$\mu$m flux of $21.8 \times 10^{-18}$~erg~s$^{-1}$~cm$^{-2}$. Table \ref{tab_gen_prop} summarizes the intrinsic properties of the two sources. 

\begin{table}[h]
%\centering
\caption{\label{tab_gen_prop} Intrinsic properties of MACS1149-JD1 and B14-65666}
\resizebox{.48\textwidth}{!}{
\begin{tabular}{ccc}
\hline
\hline
 & MACS1149-JD1$^{(1,*)}$ & B14-65666$^{(2)}$\\
\hline
\hline
Redshift & 9.11 & 7.15 \\
L$_{[OIII]88\mu m}$ [L$_{\odot}$] &  7.4$\times10^{7}$ & 34.4$\times10^{8}$\\
SFR [M$_{\odot}$~yr$^{-1}$] & 4.2 & 200 \\
M$_{*}$ [M$_{\odot}$] & 1.1$\times10^{9}$ & 7.7$\times10^{8}$ \\
sSFR [Gyr$^{-1}$]& 3.8 & 259 \\
\hline
\hline
\end{tabular}}

$^{(1)}$ \cite{Hashimoto2018b}, $^{(2)}$ \cite{Hashimoto2019}\\
$^{(*)}$ Intrinsic physical properties (after magnification correction of $\mu \sim$ 10)
\end{table} 

\subsection{Generating MRS simulated spectra}

The process to build a final calibrated 1D MRS simulated spectrum has four different phases. First, a variety of spectral templates that cover the expected range of metallicities and excitation conditions of the ionized gas for galaxies in the EoR are built (Sect. \ref{Met:Templates}).  Second, we take advantage of the MIRI instrument simulator (MIRISim)\footnote{It is part of MIRICLE python environment (\url{http://www.miricle.org})} to generate simulated MRS observations where the spectral template, astronomical scene, and instrumental and observational configurations are set up (Sect. \ref{MeT:MIRISim}). Third, the official JWST calibration pipeline is used to calibrate the simulated MRS observations and derive the 3D spectral cubes (Sect. \ref{Met:pipeline}). Finally, we extract the final 1D calibrated spectrum for each simulated MRS observation and calculate the emission line fluxes (Sect. \ref{Met:emi_flux}).

\subsubsection{Low-metallicity spectral templates}\label{Met:Templates}

\begin{table*}[h]
%\centering
\caption{\label{tab_MRS_templates}Spectral templates for MRS simulations of EoR sources.}
%%\resizebox{\textwidth/1}{!}{
%\begin{tabular}{ccccc}
%\hline
%\hline
%Template & Metallicity & [OIII]/H$\alpha$ & Source & Reference \\
%    &   (12+$\log$(O/H) & & & \\
%\hline
%\hline
%%\multicolumn{5}{c}{MACS1149-JD1}\\
%TM\_0.2\_solar & 8.02$^{(a)}$ & 1.93 & Average blue dwarfs  & %\citealt{Izotov2011} \\
%TM\_0.04\_solar & 7.29 & 1.10 & SBS0335-052E & \citealt{Izotov2011}\\
%TM\_0.02\_solar & 6.98 & 0.59 &  J0811+4130 & \citealt{Izotov2018}\\
%\hline
%\hline
%\end{tabular}%}

\resizebox{\textwidth/1}{!}{
\begin{tabular}{cccccccccccccccc}
\hline
\hline
Template & Metallicity & Source & H$\beta$ & [OIII] & H$\alpha$ & [NII] & [SII] & [SIII] & [SIII] & Pa-$\epsilon$ & Pa-$\delta$ & HeI & Pa-$\gamma$ & Pa-$\beta$  \\
 & (12+$\log$(O/H) & & 0.4861 & 0.4959 & 0.6563 & 0.6583 & 0.6716 + 0.6731 & 0.9069 & 0.9532 & 0.9550 & 1.005 & 1.087 & 1.093 & 1.282 \\
\hline
\hline
%\multicolumn{5}{c}{MACS1149-JD1}\\
TM\_0.2\_solar$^{(a)}$ & 8.02$^{(c)}$ & blue dwarfs & 0.19 & 0.33 & 0.52 & 0.018 & 0.042 & 0.021 & 0.057 & 0.006 & 0.010 & 0.068 & 0.015 & 0.026 \\
TM\_0.04\_solar$^{(a)}$ & 7.29 & SBS0335-052E & 0.33 & 0.33 & 0.91 & 0.002 & 0.014 & 0.010 & 0.022 & 0.012 & 0.016 & 0.088 & 0.024 & 0.040 \\
TM\_0.02\_solar$^{(b)}$ & 6.98 & J0811+4130 & 0.61 & 0.34 & 1.68 & 0.004 & 0.014 & --- & --- & --- & --- & 0.14$^{(d)}$ & --- & ---\\
\hline
\hline
\end{tabular}}

{Note: Line ratios are normalized to the flux of [OIII]0.5007$\mu$m emission line.\\
References: \cite{Izotov2011}$^{(a)}$ and \cite{Izotov2011}$^{(b)}$\\
$^{(c)}$ Average of O/H values from II Zw 40, Mrk 71, Mrk 930, and Mrk 996. Their metallicity range is 7.85<12+$\log$(O/H)<8.10.\\
$^{(d)}$The template TM\_0.02\_solar does not include emission lines redder than [SII]~0.6731$\mu$m because only optical spectral lines are  available. The HeI1.087$\mu$m flux have been calculated from HeI~0.5876$\mu$m, using the line ratio (HeI~1.087$\mu$m/HeI~0.5876$\mu$m) derived using the templates TM\_0.2\_solar and TM\_0.02\_solar.}

\end{table*} 

\begin{figure*}[htb!]
  \centering
%$\begin{array}{rl}
    \includegraphics[width=\hsize]{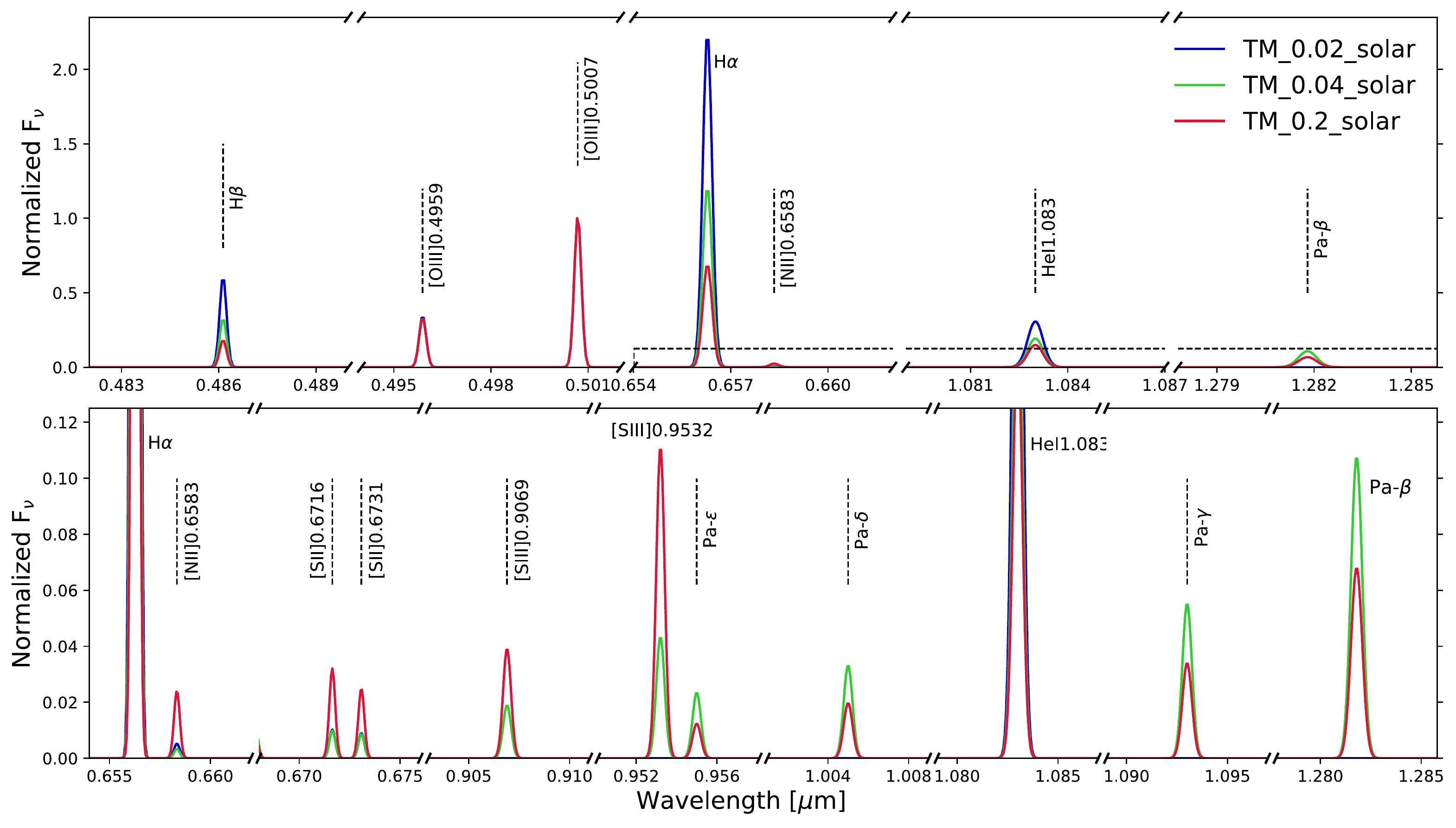}
%\end{array}$
\caption{Spectral templates used on the MRS simulated observations (see Table \ref{tab_MRS_templates} for details). Upper panel shows the brightest optical and near-IR emission lines in the range from H$\beta$ to Pa$\beta$. The bottom panel is a zoom-in of the dashed line rectangle shown in the upper panel, and illustrates the fainter optical and near-IR emission lines in the range from H$\alpha$ to Pa$\beta$. The spectra have a line width of 154~km~s$^{-1}$ (FWHM) and are normalized to  the peak of the [OIII]0.5007$\mu$m emission line; the continuum is set to zero.}\label{fig:Templates}
\end{figure*}

The spectral templates consist of only rest-frame optical/near-IR emission lines (i.e. no stellar continuum included), where the line ratios are based on observed spectra of low-$z$, low-metallicity, dwarf galaxies (Figure \ref{fig:Templates} and Table \ref{tab_MRS_templates}). To cover the range of metallicities expected in EoR sources, three different templates are constructed according to metallicity: one low-metallicity ($\sim$0.2~Z$_{\odot}$) and two metal-poor (0.04 and 0.02~Z$_{\odot}$). For the low-metallicity template (METAL\_0.2\_SOLAR) the emission lines are taken as the average ratios derived for a sample of well-measured, low-metallicity (0.2~Z$_{\odot}$), dwarf galaxies \citep{Izotov2011}. For the metal-poor templates the spectra of the metal-deficient, blue compact dwarf SBS0335-052E (\citealt{Izotov2011}, METAL\_0.04\_SOLAR), and the lowest metallicity J0811+4130 dwarf star-forming galaxy (\citealt{Izotov2018}, METAL\_0.02\_SOLAR) are used. The selected values cover the wide range of metallicities derived for FIRSTLIGHT EoR galaxies \citep{Ceverino2019}. The spectral templates do not include any contribution from a low-luminosity active galactic nucleus (AGN). In the optical range the presence of an AGN  increases the luminosity of the metallic lines relative to hydrogen, and therefore would help to detect the presence of an AGN \citep{Kewley2013}.

The templates are further normalized in flux using the R[OIII], and the observed [OIII]88$\mu$m flux.
The line ratios between the optical and far-infrared [OIII] lines have a well-known and strong dependency on the electron density and temperature of the ionized gas \citep{Dinerstein1985,Keenan1990}. According to these authors, the R[OIII] ratio has a value of $\sim$3 to $\sim$15 for an electron density of 100 cm$^{-3}$, and electron temperatures in the (1$-$2) $\times$ 10$^4$ K range. For a given temperature in this range, the R([OIII]) ratio would further increase with density by factors of up to 2 for electron densities of up to 1000 cm$^{-3}$. 
 
Studies of a large representative sample of giant HII regions, nearby HII galaxies, and green peas covering the 8.5 to 7.2 (12+$\log$[O/H])\footnote{Throughout the paper a value of 8.69 is assumed for the solar abundance, as given by \cite{Asplund2009}} metallicity range show a well-defined relation between the metallicity and the [OIII] electron temperature given by the expression \citep{Amorin2015}  12+log(O/H) = 9.22 - 0.89$\times$T$_e$([OIII], in units of 10$^4$ K). According to this expression, the electron temperature of the [OIII] ionized gas ranges from 1.4 to 2.3 $\times$ 10$^4$ K for metallicities $\sim$8.4 to $\sim$7.2  (\citealt{Amorin2015}). These temperatures agree well with those measured in a sample of high-ionization, metal-deficient, blue dwarf galaxies \citep{Thuan2005}. Blue dwarfs have [OIII] temperatures in the range of 1.3-1.4, 1.4-1.8, and 1.8-2.0 $\times$ 10$^4$ K for metallicities above 8.0, between 7.5 and 7.9, and below 7.5, respectively. On the other hand, the electron density derived from the [SII] doublet line ratio has values in the 110 to 1310 cm$^{-3}$ range, with an average of 410 cm$^{-3}$. These densities are similar to those measured in some of the most metal-deficient, low-$z$ galaxies known, such as SBS0335-052E \citep{Izotov2009}, J0811+4730 \citep{Izotov2018}, or A198691 \citep{Hirschauer2016}. In summary, the physical conditions of the [OIII] emitting gas in low-metallicity and metal-poor galaxies favour electron densities above 100 cm$^{-3}$, and temperatures well above 10$^4$ K, and closer to 2$\times$ 10$^4$K. Therefore, following the dependence of R([OIII]) with temperature and density, the spectral templates are normalized in flux with two different R([OIII]) ratios, R([OIII]) = 6.5 for the low-metallicity template (i.e. 0.2~Z$_{\odot}$)  and R([OIII])= 10 for the metal-poor template (i.e. 0.04-0.02~Z$_{\odot}$).

The width of the emission lines in the templates is simulated by a Gaussian with a full width at half maximum (FWHM) of the [OIII]88$\mu$m flux measured in each galaxy (i.e. 154~km~s$^{-1}$ for MACS1149-JD1, and 300 and 267~km~s$^{-1}$ for the components of the B14-65666 system). Finally, the templates are normalized to the [OIII]0.5007$\mu$m flux derived from the [OIII]88$\mu$m flux, and redshifted to the corresponding observed wavelengths. Galaxies at redshifts above 6 show a steep UV continuum slope ($\beta < -2$, \citealt{Bouwens2016}), in other words an optical extinction A$_V < 0.3$ mag, and therefore no internal extinction correction is applied to the line fluxes in the templates. 

Finally, the UV-brightest sources at $z > 7$ \citep{Roberts-Borsani2016,Stark2017} have continuum fluxes of 0.2-0.4~$\mu$Jy at 4.5$\mu$m. The 10$\sigma$ sensitivity for a 10ks observation with the MRS Channel 1 is $\sim$35-55~$\mu$Jy \citep{Glasse2015}, depending of the wavelength. The continuum emission is well bellow the detection limit of the MRS in the exposure time used here. Then the templates only contain the main optical and near-IR emission lines in the H$\beta$ to Pa$\beta$ spectral range without continuum emission.

\subsubsection{MIRI instrument simulator: MRS raw observations}\label{MeT:MIRISim}

We use MIRISim (Klaassen et al. in prep.), public release 2.0.0,\footnote{Public and stable MIRISim releases are available at \url{http://miri.ster.kuleuven.be/bin/view/Public/MIRISim_Public}.} to perform simulated MRS observations of the EoR sources, MACS1149-JD1 and B14-65666. MIRISim is the MIRI instrument simulator able to reproduce realistic observations with the MRS and with other MIRI observational modes. It takes advantage of the full information collected during the cryogenic test and calibration campaigns of MIRI to simulate realistic point spread function (PSF), detector read noise, Poisson noise, dark current, detector non-linearity, flat-fielding, cosmic rays, fringing, and other observational and instrumental effects. MIRISim allows modelling of astronomical targets, combining SEDs and emission line information with different morphologies, and with user-provided astronomical images. It produces the raw uncalibrated data that are input into the MIRI JWST calibration pipeline to obtain the calibrated data cube.

The lensed galaxy detected at $z=9.11$ with ALMA, MACS1149-JD1,  presents [OIII]~88$\mu$m observed flux of $3 \times 10^{-18}$~erg~s$^{-1}$~cm$^{-2}$ with a line width of 154~km~s$^{-1}$ (FWHM, \citealt{Hashimoto2018b}). Its strongest optical and near-IR lines (H$\beta$, [OIII]~0.4959,0.5007$\mu$m, H$\alpha$, and HeI~1.087$\mu$m) fall in MRS Channels 1 and 2.\footnote{MRS has wavelength ranges in Channel 1 (4.89 < $\lambda_{obs}$[$\mu$m] < 7.66) and Channel 2 (7.49 < $\lambda_{obs}$[$\mu$m] < 11.71), and its resolving power ranges are 2750 < $\lambda/\Delta\lambda$ < 3610.} In order to investigate their detectability as a function of metallicity and electron temperature and density, we simulate three medium-deep (10~ks) MRS observations with different spectral templates and R[OIII] ratios (see Table \ref{tab_MRS_sim} for details). We consider MACS1149-JD1 as an unresolved source for the MRS, and located in the centre  of the Channel 1 field of view. The solar activity, which is related with the frequency of cosmic rays events, and the instrument and sky backgrounds are set to low. A four-point dither pattern is used to generate the MRS observations. Each of the dither pointings consists of 35 groups, three integrations, and one exposure in SLOW read-out mode, which gives 2.5~ks of integration per pointing, for a total of 10ks on-source integration time per MRS spectral setting\footnote{Information about wavelength coverage, spectral setting, spatial resolution, dithering pattern, detector read-out mode, and exposure time for the MRS can be  found at \url{https://jwst-docs.stsci.edu/display/JTI/MIRI+Medium-Resolution+Spectroscopy}}. We note that a MRS spectral setting (SHORT, MEDIUM, or LONG) covers one-third of the available wavelength range in each channel; therefore, the three different spectral settings are needed for full spectral coverage. For MACS1149-JD1 simulations, we use two spectral settings, SHORT and LONG.

The interacting or merging system at redshift of $z=7.15$, B14-65666,  is composed of two UV-bright sources with a projected separation of 2-4 kpc. The system presents a total integrated [OIII]~88$\mu$m flux of $21.8 \times 10^{-18}$~erg~s$^{-1}$~cm$^{-2}$. We simulated a deep (40~ks) Channel 1 and 2 MRS observation to investigate the possibility of detecting H$\alpha$ and other weak optical and near-IR emission lines ([NII]0.6583$\mu$m, [SII]~0.6716,0.6731$\mu$m, [SIII]~0.9069,0.9532$\mu$m, Paschen series). B14-65666 is simulated combining two unresolved sources with a separation of 1". The full extension in [OIII]88$\mu$m is 0.84", and the separation between clumps in rest-frame UV is around 0.5". Since the optimal deblending of two sources in the MRS observations is beyond the scope of this paper, the separation between clumps has been increased to reduce the confusion. B14-65666\_0.2\_solar is simulated at a redshift of 7.153, with [OIII]~88$\mu$m flux of $13.5 \times 10^{-18}$~erg~s$^{-1}$~cm$^{-2}$, R[OIII] of 6.5, and a line width of $\sim$325~km~s$^{-1}$ (FWHM). B14-65666\_0.04\_solar is simulated at a redshift of 7.1482, with [OIII]~88$\mu$m flux of $8.3 \times 10^{-18}$~erg~s$^{-1}$~cm$^{-2}$, R[OIII] of 10, and a line width of $\sim$267~km~s$^{-1}$ (FWHM). We note that an offset in velocity between the two components has been included, as presented in \cite{Hashimoto2019}. A different spectral template and R[OIII] ratio is used for each component to analyse the detectability of the lines with different metallicity and physical conditions (see Table \ref{tab_MRS_sim}). The solar activity and the instrument and sky background are set to low. An eight-point dither pattern is used to generate the MRS observations. Each of the dither pointings consists of 35 groups, three integrations, and two exposures in SLOW read-out mode, which gives 5~ks of integration per pointing, and a total of 40~ks on-source integration time per MRS spectral setting (SHORT, MEDIUM, and LONG).

Epoch of Reionization sources are expected to have sizes of less than 1 kpc \citep{Shibuya2019}, and therefore are point-like sources for the MRS PSF\footnote{FWHM  $\sim$ 0.31"-0.42" depending on the Channel; see \citealt{Wells2015} for an extensive explanation of the PSF dependence with the wavelength}; galaxies at lower redshifts would be larger in size, with a median radius of 2.2 kpc \citep{Ribeiro2016}. This would imply a dilution of the observed flux over a larger number of spaxels, and therefore would require these galaxies to be treated as extended sources with a specific light profile and clumpiness in the simulations.

Alternatively, the SED-fitting SFRs could be used to derive the H$\alpha$ emission \citep{Kennicutt1998}. MACS1149-JD1 and B14-65666 system have a intrinsic SFRs of 4.2 and 200 M$_{\odot}$~yr$^{-1}$ that is equivalent to observed H$\alpha$ fluxes of 8 and 67 $\times~10^{-18}$~erg~s$^{-1}$~cm$^{-2}$, respectively. The predicted H$\alpha$ fluxes are in close agreement with the low-metallicity templates derived using the methodology presented in Sect. \ref{Met:Templates}.

\begin{table}[h]
%\centering
\caption{\label{tab_MRS_sim}Properties of the MACS1149-JD1 and B14-65666 MRS simulated templates}
\resizebox{.48\textwidth}{!}{
\begin{tabular}{ccccc}
\hline
\hline
Simulated\_Spectrum & Template & R[OIII]$^{(1)}$ & F$_{[OIII]}^{(2)}$ & F$_{H\alpha}^{(2)}$\\
\hline
\hline
MACS1149\_0.2\_solar & TM\_0.2\_solar & 6.5 & 19.5 & 10.1\\
MACS1149\_0.04\_solar & TM\_0.04\_solar & 10 & 30 & 27.3\\
MACS1149\_0.02\_solar & TM\_0.02\_solar & 10 & 30 & 50.7\\
\hline
\hline
B14-65666\_0.2\_solar & TM\_0.2\_solar & 6.5 & 87.7 & 45.5\\
B14-65666\_0.04\_solar& TM\_0.04\_solar & 10 & 83.2 & 75.6\\
\hline
\hline
\end{tabular}}

$^{(1)}$ R[OIII] = F([OIII]0.5007$\mu$m)/F([OIII]88$\mu$m)\\
$^{(2)}$ Flux given in units of $10^{-18}$~erg~s$^{-1}$~cm$^{-2}$
\end{table} 

\subsubsection{Calibration of MRS observations}\label{Met:pipeline}

\begin{figure}[htb!]
 \centering
%$\begin{array}{rl}
    \includegraphics[width=\hsize]{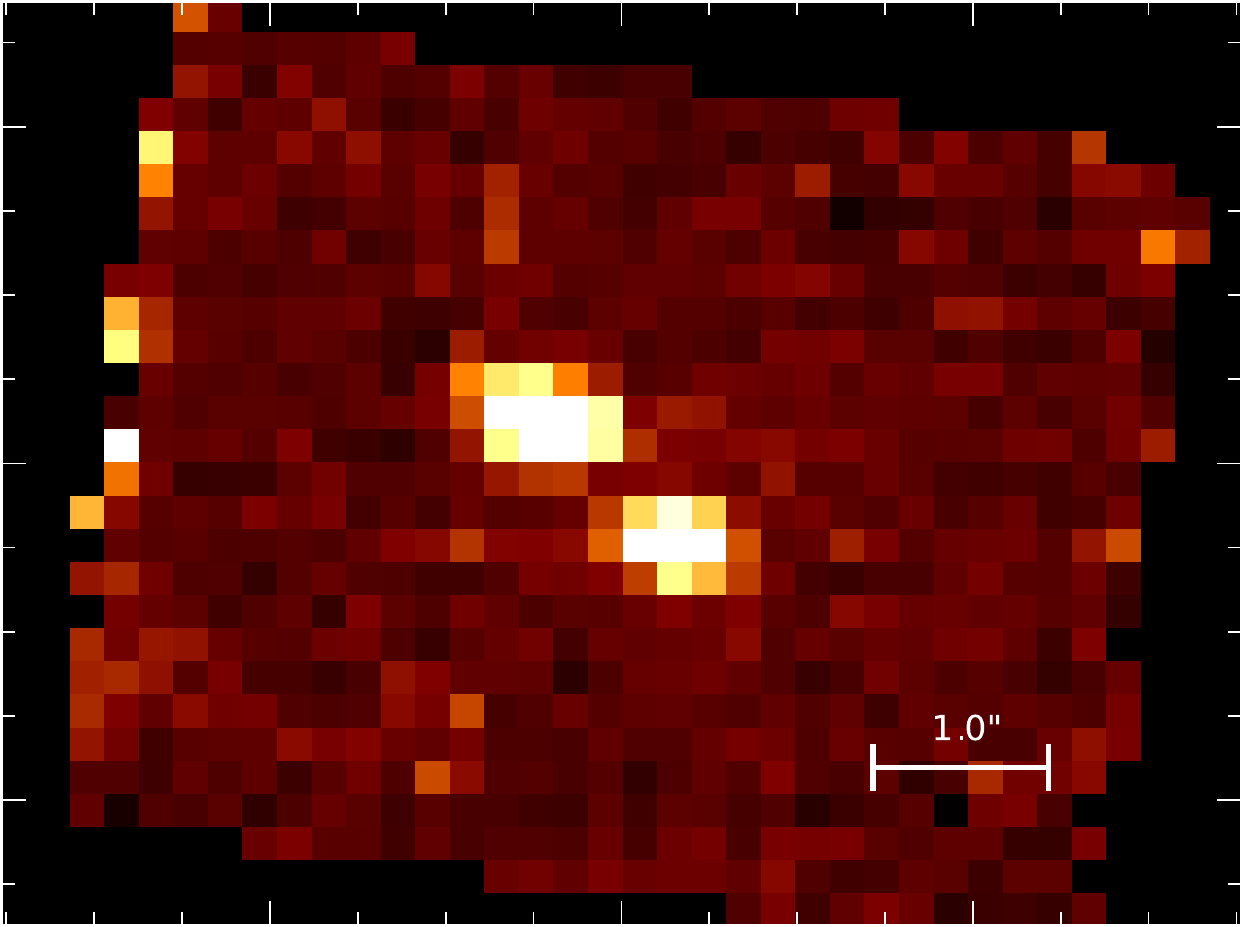}
%\end{array}$
\caption{Simulated B14-65666 system. It illustrates the integrated H$\alpha$ map for a deep (40ks) MRS observations, where two components are simulated as point-like source separated by 1", and with metallicities of 0.2 and 0.04 Z$_{\odot}$.}\label{fig:2dimageMRS}
\end{figure}

The MACS1149-JD1 and B14-65666 MIRISim simulated MRS observations are calibrated with the JWST calibration pipeline (release 0.9.6).\footnote{For more information about the JWST pipeline, see \url{https://jwst-docs.stsci.edu/jwst-data-reduction-pipeline}} The pipeline is divided into three different processing stages. The first stage performs a detector-level correction, where the MRS observations are corrected for saturation, linearity, and dark current. It also applies the jump detection and ramp-fitting modules to transform the raw MRS ramps observations to slope detector products. We use a rejection threshold of 1.75$\sigma$ to identify the jumps between adjacent frames and correct the cosmic ray events. The selected rejection threshold is optimized to produce the best S/N on the final calibrated spectrum. The modification of the rejection threshold from 4$\sigma$ to 1.75$\sigma$ produces variations of S/N in factors of $\sim$1.5 and $\sim$1.25 for Channel 1 and 2, respectively. These variations are likely to be relevant during on-orbit operations as the solar activity, and therefore changes in the density and energy of cosmic rays, could have different residual effects in the calibrated data. The second stage corrects the slope products from flat-fields  and fringes, assigns the coordinate system, and produces a photometric calibration at individual exposure levels. We note that the pipeline and MIRISim use the same reference file to simulate and calibrate the effect of the  fringes. It could underestimate the fringe residuals in the final MRS simulated spectra, which are expected to be lower than 2\%. The third stage combines the different dither exposures to create a 3D spectral cube. The final cubes have a spatial and spectral resolution of 0.196"~$\times$~0.196"~$\times$~0.001~$\mu$m for Channel 1, and 0.196"~$\times$~0.196"~$\times$~0.002~$\mu$m for Channel 2. Figure \ref{fig:2dimageMRS} shows an example of the MRS calibrated 3D spectral cubes, and illustrates the integrated H$\alpha$ map of the simulated B14-65666 system (see detailed explanations and caveats in Sect. \ref{MeT:MIRISim}).

\subsubsection{Extraction and analysis of 1D MRS spectra}\label{Met:emi_flux}

The 1D spectra are obtained by performing circular aperture photometry with a radius equal to the PSF FWHM (r $\sim$ 0.31"-0.42" depending on the Channel). The subtracted background is obtained in an annulus from 0.78" to 1.37" centred in the source. An aperture correction is applied to obtain the final 1D calibrated spectra. The aperture correction is calculated by combining simulated bright point sources on MIRISim and the PSF model obtained during the test and calibration campaigns of MIRI. The aperture correction is calculated in each wavelength of the spectral cube, Channel 1 presents values from 1.59 to 1.69 and  Channel 2 from 1.64 to 1.89.  

The emission line fluxes are derived by fitting a single Gaussian model to the line profile. The fit is performed within a spectral range equal to 0.1$\mu$m and 0.14$\mu$m for Channel 1 and 2, respectively. If the defined spectral range contains more than one emission line, we use a multiple Gaussian model to simultaneously fit the different line profiles. To estimate the flux error of each emission line, we implement a Monte Carlo method. We measure the noise of the spectra as the root mean square (rms) of the residuals after subtracting the derived Gaussian profile. The noise is used to generate N ($N=3000$) new spectra, where a random Gaussian noise with a sigma equal to the rms is added to the original spectrum and the lines are again fitted. The error of the measurements is obtained as the standard deviation of the N derived fluxes. Tables \ref{tab_fluxes_MACS} and \ref{tab_fluxes_B14} contain the derived integrated fluxes and uncertainties for the optical and near-IR emission lines analysed in the simulated MACS1149-JD1 and B14-65666 MRS observations. 

\begin{table*}[htb!]
\centering
\caption{\label{tab_fluxes_MACS} Derived emission line fluxes for the MACS1149-JD1 simulated spectra.}
%\resizebox{\textwidth/1}{!}{
\begin{tabular}{cccccc}
\hline
\hline
Simulated\_Spectrum$^{(a)}$ & H$\beta$ & [OIII] & [OIII] & H$\alpha$ & HeI\\
 & 0.4861 & 0.4959 & 0.5007 & 0.6563 & 1.087\\
\hline
\hline
MACS1149\_0.2\_solar & 3.1$\pm$0.7  & 5.4$\pm$0.7 & 18.8$\pm$0.7 & 9.2$\pm$0.5  & 2.0$\pm$0.4 \\
MACS1149\_0.04\_solar & 9.1$\pm$0.7 & 9.5$\pm$0.7 & 28.6$\pm$0.7 & 25.0$\pm$0.5 & 2.2$\pm$0.4 \\
MACS1149\_0.02\_solar & 16.7$\pm$0.7 & 9.7$\pm$0.7 & 30.0$\pm$0.7 & 44.8$\pm$0.5 & 3.9$\pm$0.4 \\
\hline
\hline
\end{tabular}%}

Note: The fluxes and noise for all emission lines and metallicities correspond to an exposure time of 10ks.\\
$^{(a)}$ flux given in units of $10^{-18}$~erg~s$^{-1}$~cm$^{-2}$
\end{table*}

\begin{figure*}[htb!]
 \centering
%$\begin{array}{rl}
    \includegraphics[width=\hsize]{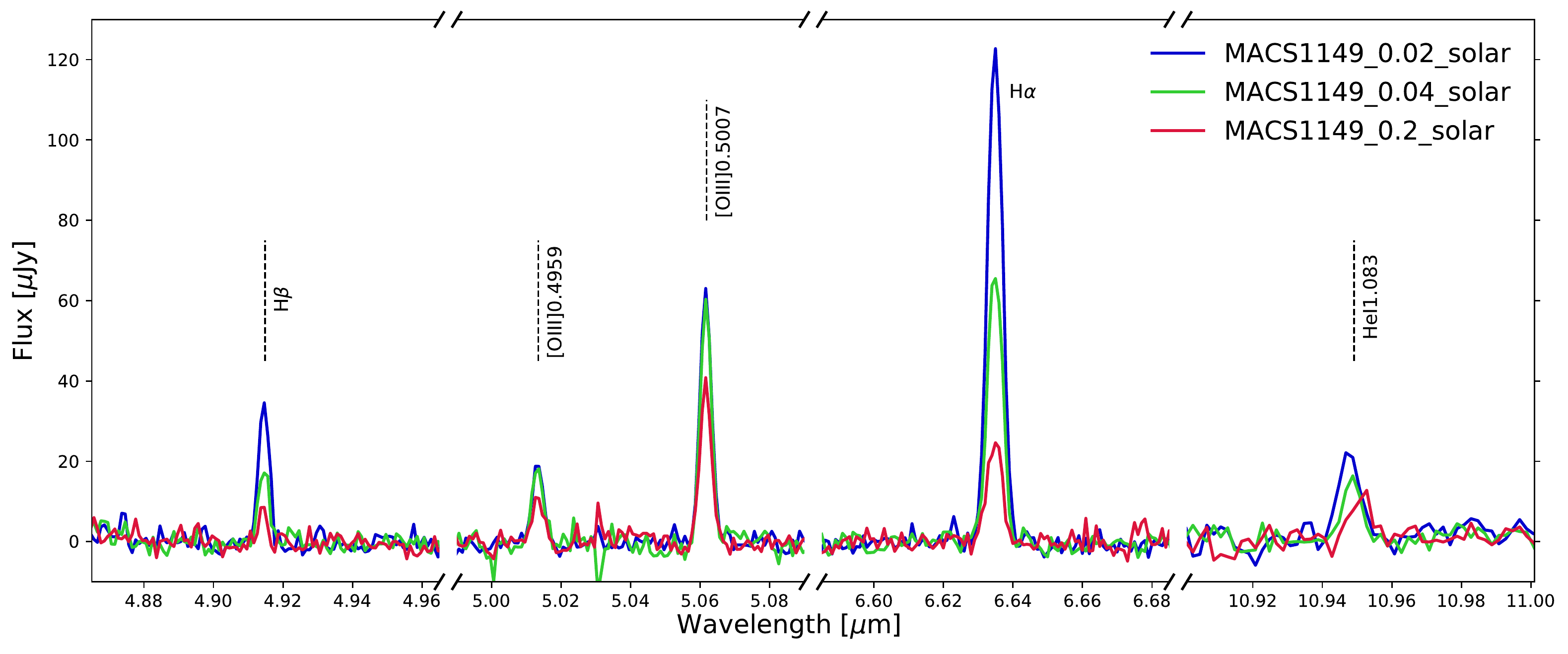}
%\end{array}$
\caption{Simulated medium-deep (10ks) MRS observation of MACS1149-JD1 at a redshift of 9.11. It illustrates the simulated spectrum with metallicities of 0.02~Z$_{\odot}$ (blue), 0.04~Z$_{\odot}$ (green), and 0.2~Z$_{\odot}$ (red). The main emission lines are shown as dashed lines, and their derived integrated fluxes can be found in Table \ref{tab_fluxes_MACS}.}\label{fig:MACSspectra}
\end{figure*}

The absolute fluxes of the emission lines detected with high significance (S/N > 10) are in agreement with the input values with average deviations lower than 10\%\footnote{The absolute photometric calibration uncertainties reported by the MIRISim team are can be found at \url{http://miri.ster.kuleuven.be/bin/view/Public/MIRISimPublicReleases}}. The same emission lines, those with S/N > 10, are also used to investigate the S/N differences between the JWST exposure time calculator (ETC)\footnote{\url{https://jwst.etc.stsci.edu/}} and the MRS simulated observations based on the combination of MIRISim and JWST pipeline. The ETC provides mean S/N values of  25\% and 6\% lower than those derived from respectively the medium-deep (10ks) and deep (40ks) MRS simulated observations for Channel 1 and 2. As we note in Sect. \ref{Met:pipeline}, the tuning of the configuration parameters of the JWST pipeline produces variations in the S/N of the final 1D spectrum. This level of difference is expected as the ETC, MIRISim, and JWST pipelines approximate our current best knowledge and understanding of the performance of MIRI, and the remaining uncertainties associated with noise properties, cosmic ray effects, and pipeline processing are still under study, and will be revised with in orbit commissioning data.

\subsection{Exploring the physical properties of EoR [OIII]88$\mu$m line emitters}

The MRS simulated 1D spectra of the [OIII]-emitters MACS1149-JD1 and B14-65666 were  analysed to investigate the detectability of their main optical and near-IR emission lines, and the prospects of inferring key physical properties such as the instantaneous star formation rates, ionization,  Ly$\alpha$ escape fractions, shape and hardness of the ionizing continuum, metallicity, among other physical properties. 

\subsubsection{EoR lensed sources: MACS1149-JD1}\label{Res:MACS}

As already mentioned in Sect. \ref{Met:FullSim} and Table \ref{tab_gen_prop}, MACS1149-JD1 is a lensed galaxy recently detected in [OIII]88$\mu$m at a redshift of 9.11. The intrinsic properties, SFR of 4.2 M$_{\odot}$ yr$^{-1}$, and sSFR of 4 Gyr$^{-1}$ place it in the upper SFR range of FIRSTLIGHT galaxies at a redshift of 9, but at the lower end of the sSFR as the total estimated stellar mass is 1.1$\times$ 10$^9$ M$_{\odot}$ \citep{Hashimoto2018b}.  Figure \ref{fig:MACSspectra} shows the simulated (10 ks) MRS spectra for a MACS1149-JD1-like source using three different metallicities (0.2, 0.04, and 0.02~Z$_{\odot}$) and R[OIII] values, covering the expected range of metallicities and excitation conditions in the ionized gas at a redshift of 9.11. The 1D extracted spectra containing the brightest optical emission lines (H$\beta$, [OIII]0.4959,0.5007$\mu$m, H$\alpha$, and HeI1.087$\mu$m) show the detection of all lines at a significance level higher than 4$\sigma$ at different metallicities. In particular, we obtain S/N $\sim$ 5-24, 8-42, 18-90, and 5-10 for the integrated fluxes of H$\beta$, [OIII]0.4959,0.5007$\mu$m, H$\alpha$, and HeI1.083$\mu$m emission lines, respectively. 

 \begin{table*}[htb!]

%\centering
\caption{\label{tab_fluxes_B14} Derived emission line fluxes for the simulated spectra of the B14-65666 system.}
\resizebox{\textwidth/1}{!}{
\begin{tabular}{ccccccccccc}
\hline
\hline
Simulated\_Spectrum$^{(a)}$ & H$\alpha$ & [NII] & [SII] & [SIII] & [SIII] & Pa-$\epsilon$ & Pa-$\delta$ & HeI & Pa-$\gamma$ & Pa-$\beta$\\
 & 0.6563 & 0.6583 & 0.6716 + 0.6731 & 0.9069 & 0.9532 & 0.9550 & 1.005 & 1.087 & 1.093 & 1.282 \\
\hline
\hline
B14-65666\_0.2\_solar & 46.1$\pm$0.6 & 1.1$\pm$0.4 & 4.7$\pm$0.6  & 1.8$\pm$0.3 & 5.0$\pm$0.3 & <0.9~$^{(b)}$ & 0.9$\pm$0.2 & 5.7$\pm$0.3 & 1.1$\pm$0.3 & 2.6$\pm$0.3\\
B14-65666\_0.04\_solar & 77.0$\pm$0.6 & <1.2~$^{(b)}$ & <1.8~$^{(b)}$ & 0.8$\pm$0.3 & 1.9$\pm$0.3 & 1.3$\pm$0.3 & 1.4$\pm$0.2 & 6.9$\pm$0.3 & 2.2$\pm$0.3 & 3.5$\pm$0.3 \\
\hline
\hline
\end{tabular}}

Note: The fluxes and noise for all emission lines and metallicities correspond to an exposure time of 40ks.\\
$^{(a)}$ flux given in units of $10^{-18}$~erg~s$^{-1}$~cm$^{-2}$\\
$^{(b)}$ 3$\sigma$ upper-limits.
\end{table*} 

\begin{figure*}[htb!]
  \centering
%$\begin{array}{rl}
    \includegraphics[width=\hsize]{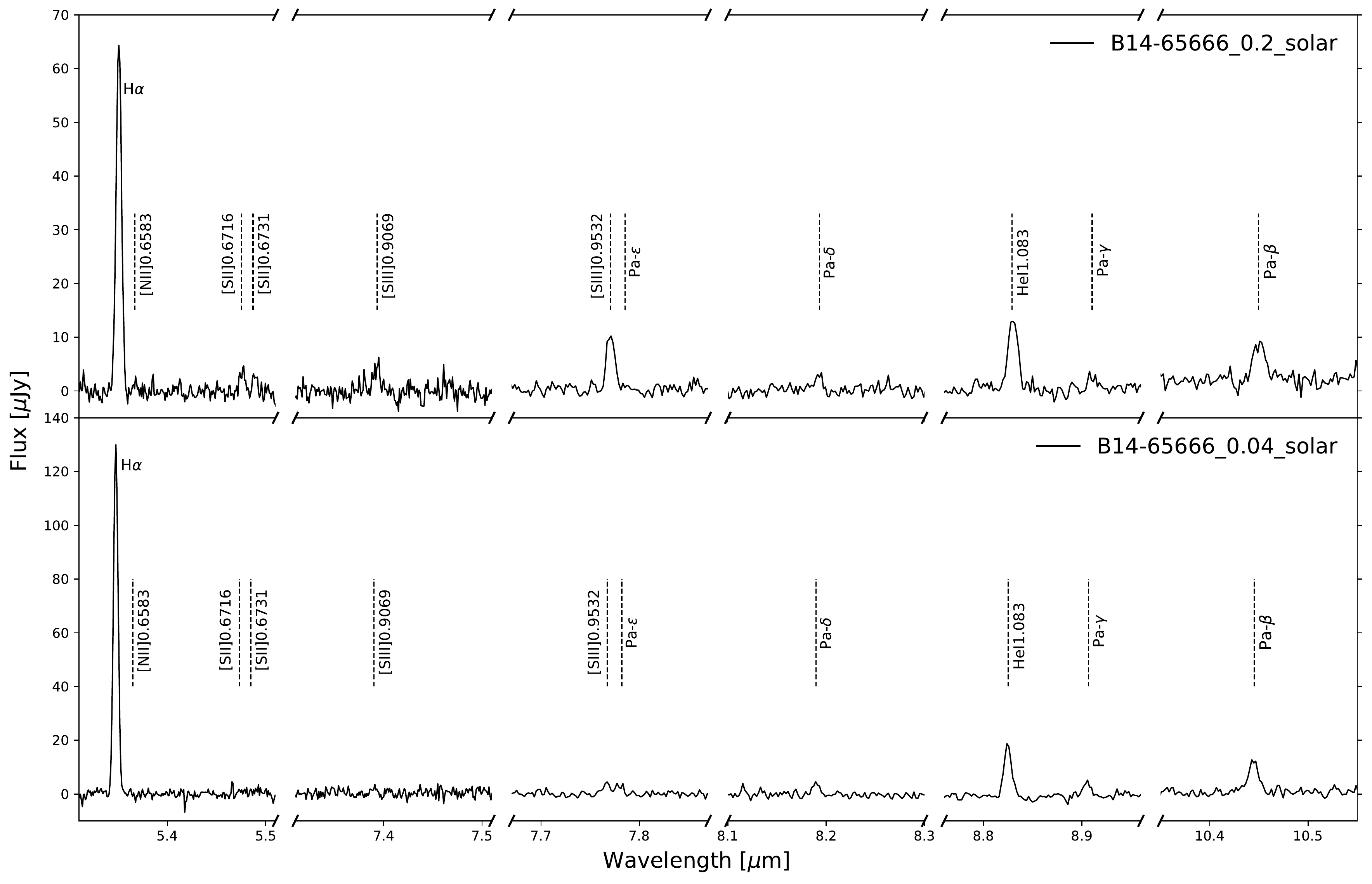}
%\end{array}$
\caption{Simulated deep (40 ks) MRS observation of B14-65666 at a redshift of 7.15. The upper panel shows the spectrum for one of the components assuming a metallicity 0.2~Z$_{\odot}$ (B14-65666\_0.2\_solar), and the bottom panel for the second component with a lower metallicity of 0.04~Z$_{\odot}$ (B14-65666\_0.04\_solar). The emission lines are shown as  a dashed line, and their derived integrated  fluxes can be found in Table \ref{tab_fluxes_B14}. }\label{fig:TDspectra}
\end{figure*}

Thanks to the additional magnification factor due to lensing, the MRS spectra illustrate important results for strong [OIII] 88$\mu$m line emitters at the highest redshifts (i.e. z $>9$). First, for a given [OIII] 88 $\mu$m luminosity, the [OIII]0.5007$\mu$m and H$\alpha$ lines will be most luminous for the lowest metallicity, and therefore would be detected with the highest significance at 0.02Z$_{\odot}$. This effect is mainly due to the expected increase in the electron temperature of the ionized gas, and therefore the R[OIII] decreases with  metallicity from  subsolar to metal poor (see Sect. \ref{Met:Templates}). Second, additional detection of the H$\beta$ emission line provides the opportunity to set direct quantitative constraints in key physical aspects of these galaxies like the internal extinction (H$\alpha$/H$\beta$ ratio) and the total instantaneous star formation rate (H$\alpha$). Third, the detection of both H$\alpha$ and HeI1.083$\mu$m, the strongest HeI in the entire UV to near-IR spectral range,  will provide unique information on the hardness of the ionizing source, even for the lowest metallicity sources. The ratio of ionizing photons can be derived as
\begin{equation}
\dfrac{Nph~[>13.6eV]}{Nph~[>24.6eV]} = (0.89-2.10) \times \dfrac{L~[H\alpha]}{L~[HeI1.083\mu m]}
\end{equation}
after extinction correction, and assuming emissivities for hydrogen \citep{Osterbrock1989book}, and HeI \citep{Benjamin1999,Porter2005} for electron densities of 100 cm$^{-3}$ and temperatures of 1-2 $\times$ 10$^4$ K, similar to those measured in low-metallicity, low-z galaxies \citep{Izotov2014}. However, as the HeI 1.083$\mu$m emissivity has a strong dependence with the electron density (factors 6 to 8 for densities in the $10^2 - 10^4$ cm$^{-3}$ range), a measure of the electron density (e.g. [SII]0.6717+0.6731$\mu$m optical lines or [OIII]52,88$\mu$m far-IR lines) is required to get an accurate value for the hardness of the ionizing source. For an instantaneous starburst and a given metallicity, the ratio of ionizing photons is a strong function of age with a drop in photons with energies above 24.6 eV relative to 13.6 eV for ages older than 6 Myr. Binaries \citep{Eldridge2017} could also be playing a relevant role in changing the hardness of the ionizing spectrum in these galaxies, in particular at low metallicities. The presence of a low-luminosity AGN could also produce an ionizing spectrum harder than the predicted from stars only. 

In addition, if Ly$\alpha$ measurements are available, the MIRI-MRS H$\alpha$ observed flux and the H$\alpha$/H$\beta$ derived internal extinction measurement, will provide a measurement of the Ly$\alpha$ escaping fraction,
\begin{equation}
F_{esc}[Ly\alpha]= \dfrac{F_{obs}[Ly\alpha]}{F_{int}[Ly\alpha]}=\dfrac{F_{obs}[Ly\alpha]}{R(Ly\alpha,H\alpha) \times F_{int}[H\alpha]},
\end{equation}
where F$_{obs}$[Ly$\alpha$] is the observed Ly$\alpha$ flux (at $z>6$),\footnote{F$_{obs}$[Ly$\alpha$]  also depends on the intergalactic medium (IGM) transmission, so that F$_{obs}$[Ly$\alpha$]=F$_{em}$[Ly$\alpha$]$\times$T$^{\rm IGM}_{\rm Ly \alpha}$, where F$_{em}$[Ly$\alpha$] is the emitted Ly$\alpha$ flux and T$^{\rm IGM}_{\rm Ly \alpha}$ is the IGM transmission to Ly$\alpha$ photons (e.g.  \citealt{Inoue2014})} F$_{int}$[H$\alpha$] is the intrinsic H$\alpha$ emission after correction for internal extinction, and R(Ly$\alpha$,H$\alpha$) is the theoretical recombination case B value assumed to be 8.7 for the typical electron densities (a few $\times$ 10$^{2}$ cm$^{-3}$ and temperatures $< 2 \times 10^4$~K). Likewise, as the H$\alpha$ line is the least affected by extinction of all the optical hydrogen recombination lines, it provides a more accurate estimate of the escape fraction of ionizing photons (e.g. \citealt{Matthee2017}) when combined with the observed rest-frame $<$912$\AA$ photometry from existing HST or future NIRCam/JWST imaging.

Finally, the high S/N of [OIII]0.5007$\mu$m and H$\alpha$ emission lines open the possibility of detecting the presence of ionized gas outflows. Although beyond  the scope of the present paper, preliminary simulations show that massive ionized outflows ($>$ 10$^7$ M$_{\odot}$, blueshifted by $\sim$300 kms$^{-1}$, and with terminal velocities of 650-700 km s$^{-1}$) could by traced by the H$\alpha$ line in metal-poor sources similar to MACS1149-JD1 (Colina et al. in prep.)

\subsubsection{UV-bright and massive EoR sources: B14-65666}\label{Res:B14}

B14-65666, as already mentioned in Sect. \ref{Met:FullSim} and Table \ref{tab_gen_prop}, is a strong [OIII]88$\mu$m line emitter at a redshift of 7.15 detected with ALMA also as a [CII]158$\mu$m source \citep{Hashimoto2019}. This UV-bright source (M$_{UV} \sim -22.3$), is identified with a system of two galaxies, likely interacting or merging. Its derived global properties with a total SFR of 200 M$_{\odot}$ yr$^{-1}$, a stellar mass of 7.7 $\times$ 10$^8$ M$_{\odot}$, a low visual extinction (A$_V$=0.3 mag),  and sSFR of 259 Gyr$^{-1}$ place it among the most massive star-forming galaxies known at a redshift above 7, excluding quasi-stellar objects (QSOs). As such, it provides with an extraordinary opportunity for the detection of faint metallic lines ([NII]0.6584$\mu$m, [SII]0.6717,0.6731$\mu$m, and [SIII]0.9069,0.9532$\mu$m), and therefore establishes strong constraints on the metallicity of the ionized gas in addition to the physical properties already mentioned in Sect. \ref{Res:MACS}. Figure \ref{fig:TDspectra} shows the deep (40 ks) MRS simulated spectra of B14-65666 assuming, for the purpose of this simulation, that one of the components of the system has a metallicity of 0.2~Z$_{\odot}$ (upper panel), while the metallicity for the second component is  0.04~Z$_{\odot}$ (bottom panel). For the 0.2~Z$_{\odot}$ spectrum, the weak [NII]0.6584$\mu$m and [SII]0.6717,0.6731$\mu$m integrated emission lines are detected at about the  3$\sigma$ level, while the [SIII]0.9069,0.9532$\mu$m integrated lines are detected at 6$\sigma$ and 17$\sigma$, respectively. On the other hand, the 0.04~Z$_{\odot}$ spectrum shows no detection (at the 3$\sigma$ level) of   [NII]0.6584$\mu$m or [SII]0.6717,0.6731$\mu$m, while the [SIII]0.9069,0.9532$\mu$m lines are detected at  3$\sigma$ and 7$\sigma$, respectively. Thus, the metallicity of luminous [OIII]88$\mu$m emitters detected by ALMA or other JWST instruments (e.g. NIRSpec) could be explored in full, using the standard R23, and all the different optical tracers as well, including the N2, S2, N2S2H$\alpha$, N2S2 ratios, as well as the combined O3N2, O3S2, and S23 ratios (see \citealt{Maiolino2019} for a review).

%--------------------------------------------------------------------
\section{Conclusions}\label{Conc}

This paper has presented a study of the H$\alpha$ fluxes predicted by state-of-the-art FIRSTLIGHT cosmological simulations for galaxies at redshifts of 6.5 to 10.5, covering the Epoch of Reionization, and of its detectability with the Medium Resolution Spectrograph (MRS) of the mid-IR Instrument (MIRI) on JWST. The paper has investigated the MRS detectability of the FIRSTLIGHT sources as a function of redshift, star formation rate, stellar mass, and specific star formation. In addition, it has presented realistic MRS simulated observation of the rest-frame optical and near-IR spectra of EoR sources recently detected by ALMA as [OIII]88$\mu$m emitters. These include the lensed source MACS1149-JD1 and the interacting-merger candidate B14-65666 at a redshift of 9.11 and 7.15, respectively. These simulations cover different metallicities and emission line ratios, and are based on medium-deep (10ks) and deep (40ks) MRS observations using the current versions of the MIRI instrument simulator (MIRISim), and of the official JWST calibration pipeline. The main conclusions are as follows:   

\begin{enumerate}
    \item All currently ALMA detected [OIII]88$\mu$m emitters at redshifts above 7 can be detected in the H$\alpha$ line with MRS spectroscopy in a few hours (10 ks) with a high significance (i.e. with S/N $>$ 5$\sigma$).
    \item Deep integrations (40 ksec) with MRS will detect (at least at the 5$\sigma$ level) H$\alpha$ emission line in EoR sources at redshifts above 7 with a SFR above $\sim$ 2 M$_{\odot}$~yr$^{-1}$, stellar masses above $\sim$ 4-9 $\times$ 10$^{7}$ M$_{\odot}$, and specific star formation above 4 Gyr$^{-1}$. 
    These limits cover the upper end of the SFR and stellar mass distribution at those redshifts, representing $\sim$ 6\% and $\sim$1\% of the predicted FIRSTLIGHT population in the 6.5-7.5 and 7.5-8.5 redshift ranges, respectively. 
    \item The FIRSTLIGHT population is dominated by numerous low-mass galaxies with faint H$\alpha$ emission for JWST spectroscopy, characterized by median values of $\widetilde{F} (\rm H\alpha) = 3.8$, $2.4$, and $1.5 \times 10^{-20}$~erg~s$^{-1}$~cm$^{-2}$ in the redshift intervals of 6.5-7.5, 7.5-8.5, and $z > 8.5$, respectively. However,  a fraction of galaxies show much higher fluxes around $F (\rm H\alpha) \sim 10^{-18}$ - $10^{-17}$~erg~s$^{-1}$~cm$^{-2}$ and are accessible to observation with MIRI/JWST spectroscopy.
    \item The MRS will provide a good S/N H$\beta$ (5-24$\sigma$) - H$\alpha$ (18-90$\sigma$) emission line spectra of sources similar to the MACS1149-JD1 at a redshift of 9.11 in exposures of a few hours ($\sim$ 10ks) for metallicity 0.2-0.02~Z$_{\odot}$. This example clearly illustrates  the possibility of performing detailed studies of intrinsically bright or lensed sources, even at the beginning of the Epoch of Reionization.
    \item The MRS will be able to establish and put strong limits on the metallicity of bright EoR sources, as demonstrated by the simulated B14-65666 system at 7.15 with metallicities 0.2 and 0.04~Z$_{\odot}$. This will be achieved by adding the optical metallicity tracers (N2, S2, N2S2H$\alpha,$ and N2S2) to the standard R23.
    \item A measure of the hardness of the ionizing spectrum, Nph(>912$\AA$)/Nph(>504$\AA$), can be derived directly from the L(H$\alpha$)/L(HeI1.083$\mu$m) line ratio if the electron density is known. This measure of the hardness will constrain the nature of the ionization source, i.e. the age and IMF upper mass limit of the stellar population, or the presence of a low luminosity AGN.
\end{enumerate}

As shown in this paper, the prospects of detecting the H$\alpha$ emission line with very high S/N (>50)  at least in bright (intrinsic or lensed) sources at redshifts of 7 to 9, opens the opportunity of investigating the presence and properties of outflows of ionized gas in galaxies during the Epoch of Reionization. 
%--------------------------------------------------------------------
\begin{acknowledgements}
The authors gratefully thank  the Referee for the constructive comments and recommendations that  helped to improve the quality of the paper, and the EC MIRI test team and MIRISim developers for providing a great and useful tool, the MIRI instrument simulator (MIRISim). The authors also acknowledge  the STScI and the developer team  of the official JWST calibration pipeline. This work was supported by the Spanish Ministry for Science, Innovation and Universities project number ESP2017-83197. D.C. acknowledges the Gauss Center for Supercomputing for funding this project by providing computing time on the GCS Supercomputer SuperMUC at Leibniz Supercomputing Centre (Project ID: pr92za). D.C. is supported by the state of Baden-W\"urttemberg through bwHPC. D.C. is a DAWN fellow. A.L. acknowledges funding from the Comunidad de Madrid, Spain, under Atracci\'on de Talento Investigador Grant 2017-T1/TIC-5213. J.P.P. and T.V.T. acknowledge financial support from UK Space Agency grants. A.A.-H. acknowledges support from the Spanish Ministry of Science, Innovation and Universities through grants AYA2015-64346-C2-1-P and PGC2018-094671-B-I00, which were party funded by the FEDER program and from CSIC grant PIE201650E36. K.I.C. acknowledges funding from the European Research Council through the award of the Consolidator Grant ID 681627-BUILDUP. 
\end{acknowledgements}

\bibliographystyle{aa}
\bibliography{Bibliography}

\end{document}